\documentclass[12pt]{article}
\usepackage{epsfig}

\def\beq{\begin{equation}}
\def\eeq#1{\label{#1}\end{equation}}
\def\eeqn{\end{equation}}

\def\beqa{\begin{eqnarray}}
\def\eeqa#1{\label{#1}\end{eqnarray}}
\def\eeqan{\end{eqnarray}}

\let\bar=\overbar

\def\Dslash{\not{\hbox{\kern-4pt $D$}}}
\def\dslash{\not{\hbox{\kern-2pt $\del$}}}

\def\msb{{\bar{\ssstyle M \kern -1pt S}}}

\def\Title#1{\begin{center} {\Large \bf{ #1} } \end{center}}

\def\TeV{TeV/$c^2$}

\def\ifb{fb$^{-1}$}
\def\V0{V$^0$}
\def\K0{K$^0$}
\def\tWs{$ t \rightarrow W s $}
\def\tWb{$ t \rightarrow W b $}
\def\tbWsb{$ \bar{t} \rightarrow W \bar{s} $}

\begin{document}

\Title{Prospects of Direct Determination of $|V_{tq}|$ CKM Matrix Elements at the LHC}

\bigskip 


\begin{raggedright}  
{\it Dr. Theodota Lagouri\index{Lagouri, Th.}\\
Department of Physics, 
Yale University, 
New Haven, CT, 
USA}
\bigskip 
\end{raggedright}

{\em Proceedings of CKM 2012, the 7th International Workshop on the CKM Unitarity Triangle,
University of Cincinnati, USA, 28 September - 2 October 2012 }

\section{Abstract}

The prospects of measuring the CKM matrix elements $|V_{tq}|$ with top 
quarks decays at the LHC are discussed here, with the top quarks produced in the
processes $pp \rightarrow t \bar{t} X$ and $pp \rightarrow t/\bar{t} X$, 
and the subsequent decays \tWs\ and/or \tbWsb.
As for the direct measurement of $|V_{tb}|$, there is also a lot of interest 
in the direct measurements of $|V_{ts}|$ and $|V_{td}|$, as the absolute values of these CKM 
matrix elements can be modified by approximately a factor 2 from their SM values.
Direct determination of these matrix elements will require a good tagging of the 
$t \rightarrow s$ transition (for $|V_{ts}|$) and $t \rightarrow d$ transition (for $|V_{td}|$) in the top quark decays, 
and a very large top quark statistics, which is available at the LHC.
Lacking a good tagging for the $t \rightarrow d$ transition, and also because of the small size of 
the CKM-matrix element, $|V_{td}| = O(10^{-2})$,  direct 
measurements of $|V_{ts}|$ at the LHC with main emphasis at the centre of mass energy 
$\sqrt{s}$ of 14 \TeV\ based on the PLB paper of Ali et al. \cite{ref2} are shown. Alternative methods 
for direct $|V_{tq}|$ determination at the LHC are also reported.

\section{Introduction}

Since the discovery of the top quark at Tevatron a lot of precise measurements have been 
undertaken at the two Fermilab experiments, CDF and D0 but also at the two LHC 
experiments ATLAS and CMS (see references in \cite{ref2}). Among the highlights are the measurements of the 
top quark mass, currently having an accuracy of better than 1\% \cite{ref5}, the $t\bar{t}$ 
production cross section with about 5.5 \% accuracy \cite{ref6}, and the observation of 
the electroweak single top production \cite{ref7, ref8}.

The cross section $\sigma ( p\bar{p} \rightarrow t/\bar{t} + X)$ has provided the first direct measurement 
of the dominant CKM-matrix element $|V_{tb}|$. It is assumed that the CKM matrix elements $|V_{td}|$ 
and $|V_{ts}|$ are much smaller than $|V_{tb}|$, but no assumption is made about the unitarity of the 
$3 \times 3$ CKM matrix. To obtain $|V_{tb}|^2$, the measured cross section is divided by the theoretical 
cross section for $|V_{tb}| = 1$. A direct combined CDF and D0 measurement of $|V_{tb}|$ using
$\sigma (p\bar{p} \rightarrow t/\bar{t} + X) = 3.14 (3.46) $ pb gave $|V_{tb}|=0.91(0.88) \pm 0.08(0.07) $ with 
$|V_{tb}|> 0.79 (0.77)$ at 95\% C.L. 

Direct determination of $|V_{tb}|$ with the experiments at the LHC is expected to reach an accuracy 
of a few percent. The determination of $|V_{tb}|$ with such an accuracy will be also very valuable to 
constrain beyond-the-SM physics models. In order to measure directly $|V_{ts}|$ at the LHC, developing efficient 
discriminants to suppress the dominant decay \tWb\ is needed.

\section{Results and Discussion}

A necessary first step in the analysis is the tagging of the top events in 
which  the $W^\pm$ decay leptonically in order to reduce the jet activity in top quark decays. 
The emerging s-quark from the top quark decay \tWs\, and the collinear gluons, which 
are present in the fragmentation process, will form a hadron jet. Then tagging on the \V0 
($K^0$ and $\Lambda$) in this jet, and measure their energy and transverse momentum 
distributions will give extra handles to suppress the dominant \tWb\ background.

Energetic \V0s are also present in the b-quark jets initiated by the decay \tWb\ and 
the subsequent weak decays $b \rightarrow c \rightarrow s$. However, in this case, the \V0s will be softer, 
they will have displaced vertexes (from the interaction point) and they will be often 
accompanied with energetic charged leptons due to the decays $b \rightarrow \ell^\pm X$. Absence 
of a secondary vertex and paucity of the energetic charged leptons in the jet provide a 
strong discrimination on the decays \tWb\ without essentially compromising the 
decays $t \rightarrow W s$. Thus, the scaled energy and transverse momentum distributions of 
the $K^0$s, $\Lambda$ and $\ell^\pm$, and the secondary decay vertex distributions 
($dN/dr$) are the quantities of principal interest.

The distance $r$ traversed in the transverse plane (i.e. the plane 
perpendicular to the beam axis) by the b-quark before decaying, 
is smeared with a Gaussian resolution to take into account realistic experimental conditions. Two 
representative r.m.s. values have been assumed, $\sigma$ (vertex) =1 mm and 2 mm. 

Detailed simulations are done using PYTHIA \cite{pythia} to model the production processes, 
gluon radiation, fragmentation and decay chains, and the underlying events. A total of 1 million 
events were generated with PYTHIA 6.4 with $|V_{ts}| = |V_{tb}| = 0.5$

Having generated these distributions, characterising the signal \tWs\ and the background 
\tWb\ events, a technique is used called the Boosted Decision Tree (BDT), a classification 
model used widely in data mining \cite{ref10, ref11} Ð to develop an identifier optimised for the 
\tWs\ decays. Both BDT and a variant of it called BDTD (here D stands for de-correlated) are used, to 
discriminate the signal events from the large backgrounds. Briefly, the generated input is used for 
the purpose of training and testing the samples. The input is provided in terms of the variables 
discussed earlier for the signal (\tWs) and the background (\tWb). This information is used to develop the splitting criteria to 
determine the best partitions of the data into signal and background to build up a decision 
tree (DT). 

The signal (\tWs) efficiencies are calculated for two cases called $bb/bs$ and $bs/ss$ 
for an assumed (Gaussian) vertex smearing with an r.m.s. value of 
2 mm and 1 mm, respectively. Concentrating on the $bb/bs$ case, when only one of the top 
(or antitop) quark decays via $t \rightarrow W^+ s$ these efficiencies lie typically between 5 \% 
(for the 2 mm smearing) and 20 \% (for the 1mm case) for a background (\tWb) rejection 
of a factor $10^3$.

\begin{figure}[htb]
\begin{center}
\epsfig{file=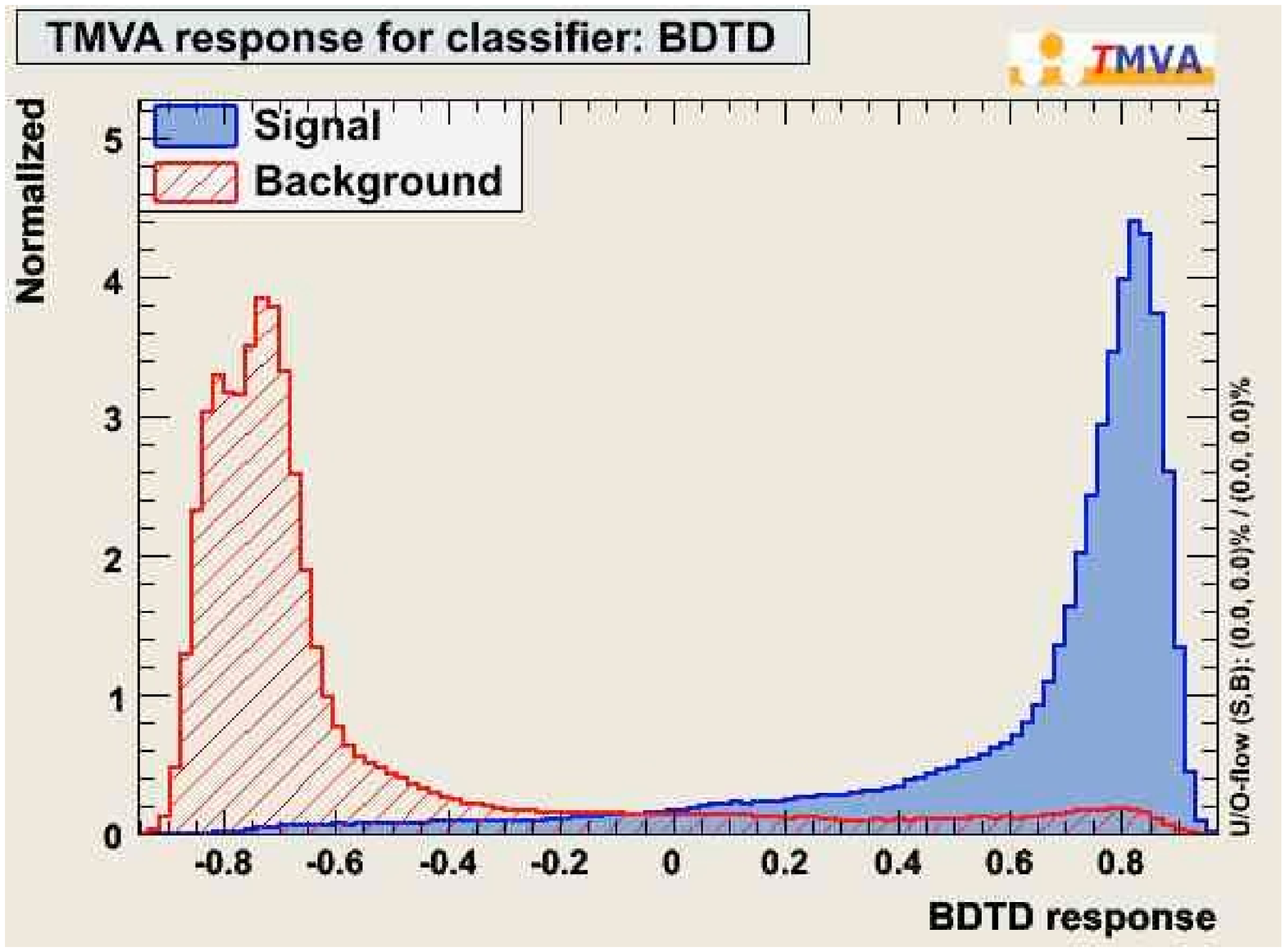,width=7cm} 
\epsfig{file=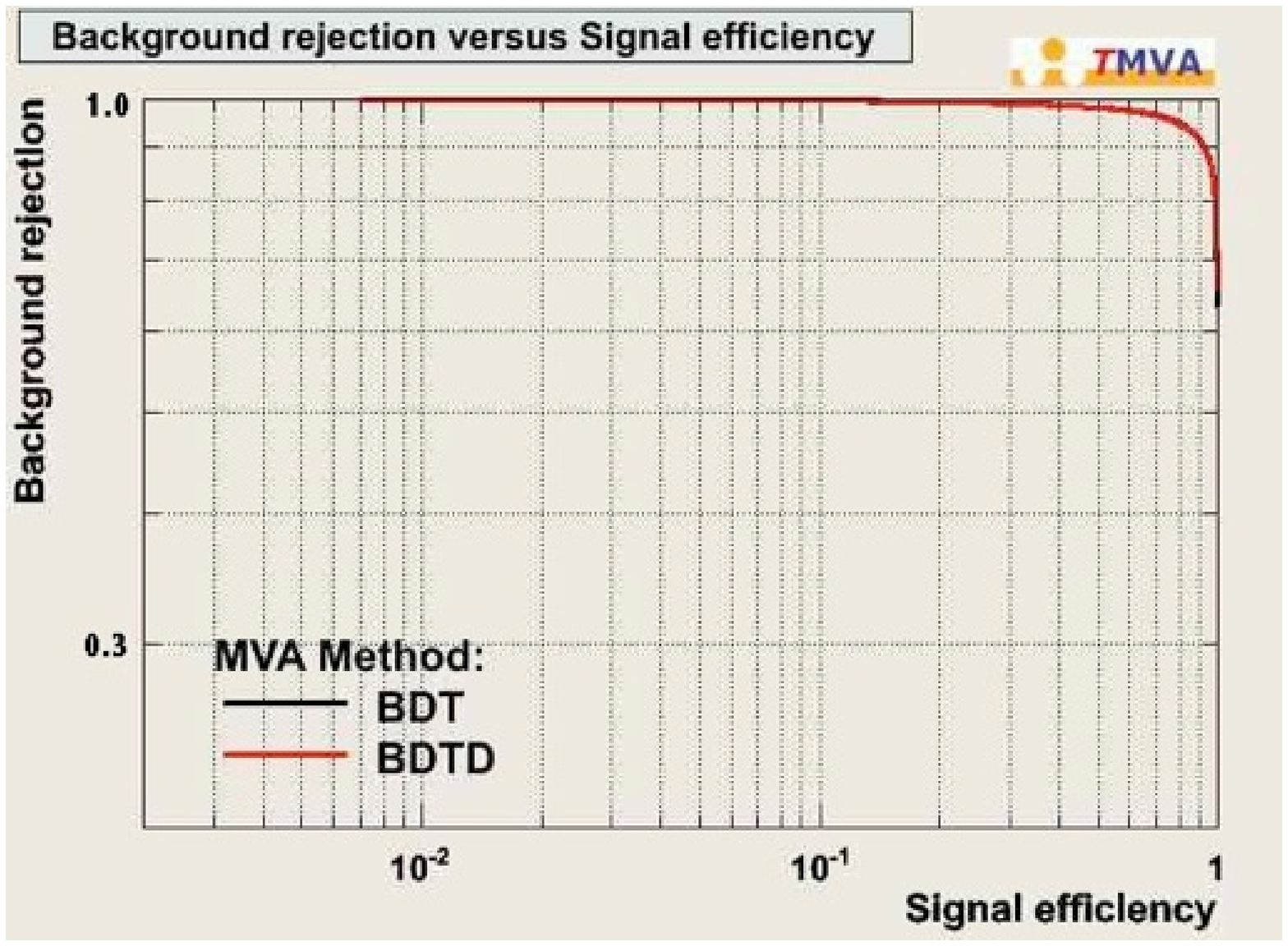,width=7cm} 
\caption{ $pp\rightarrow t \bar{t} + X$ at $\sqrt{s} = 14$ \TeV. Left frame: The normalised BDTD response. 
The signal (dark shaded) from the decay \tWs\ 
and the background (light shaded with dotted lines) from the decay \tWb\ are clearly 
separated. Right frame: Background rejection vs. signal efficiency calculated from the BDT(D) response.}
\label{fig:fig1}
\end{center}
\end{figure}

In Figure~\ref{fig:fig1} (left frame), the BDTD response functions are shown, showing that a clear separation between 
the signal (\tWs) and background (\tWb) events has been achieved. 
The background rejection vs. signal efficiency for the $pp \rightarrow t\bar{t}$ events is shown in
the right frame of Figure~\ref{fig:fig1} for both the BDT and BDTD classifiers, which give very similar results.

This level of background rejection is necessary due to the anticipated value of the ratio
$|V_{ts}|^2 / |V_{tb}|^2 \simeq 1.6 \times 10^{-3}$. The required integrated LHC luminosity to 
determine $|V_{ts}|$ directly is estimated to be 10 \ifb\ at 14 \TeV. Numerical analysis has been 
carried out for three representative LHC energies: 7 \TeV, 10 \TeV\ and 14 \TeV, here 
only the results for 14 \TeV\ are mentioned, as the distributions for 7 \TeV\ and 
10 \TeV\ are similar to the 14 \TeV\ case.

\section{Alternative methods for direct $|V_{tq}|$ measurements at the LHC}

Alternative methods of determining the matrix elements $|V_{td}|$, $|V_{ts}|$ and $|V_{tb}|$ at the LHC 
are based on the single top (or anti-top) production at the LHC. One attempts to determine these 
matrix elements from the cross section measurement by a simultaneous fit \cite{ref14}. 
In the SM, one 
expects $|V_{ts}|^2 / |V_{tb}|^2 \sim 1.6 \times 10^{-3}$ and $|V_{td}|^2 / |V_{tb}|^2 \sim 6 \times 10^{-5}$. In the 
example of realistic beyond-the-SM physics, these CKM matrix element 
ratios could be larger by a factor 4. Both in the SM, and in the four generation extension 
of it, the cross sections $\sigma (pp \rightarrow tX)$ and $\sigma(pp \rightarrow \bar{t} X)$ are completely 
dominated by the $|V_{tb}|^2$ term. Hence, this proposal does not have the desired sensitivity to 
measure the matrix elements $|V_{td}|$ and $|V_{ts}|$ at the level of theoretical interest.

It has been recently suggested \cite{ref15} that one may improve the sensitivity to $|V_{td}|$, 
using the top quark rapidity distributions, which are different for 
the valence d-quark initiated processes as opposed to the sea b-quark initiated processes. 
The described method based on the top quark decay characteristics to determine $|V_{ts}|$ complements the existing proposal.

The ratio of the CKM matrix elements $(|V_{td}|^2 + |V_{ts}|^2)/|V_{tb}|^2$, that can be obtained by measuring 
the ratio $R=B(t\rightarrow Wb)/B(t\rightarrow Wq)$, where $q=b,s,d$, through the number 
of events with zero-, one-, and two b-tags in the process $pp \rightarrow t/\bar{t} X$ \cite{ref14} can 
be combined with the determination of the ratio $|V_{ts}|^2 / |V_{tb}|^2$ discussed before, to constrain (or 
measure) the quantity $|V_{td}|^2 / |V_{tb}|^2$.

Finally, a model-independent extraction of $|V_{tq}|$ matrix elements from top-quark measurements 
\cite{ref16} is triggered by a recent D0 measurement of the ratio $R = 0.90$. 
It allows to extract the quark mixing matrix elements $|V_{td}|$, $|V_{ts}|$, and $|V_{tb}|$ 
from the measurement of $R$ and from single-top event yields. This method provides information that can be directly used to 
put constraints on the four-family extended SM and other scenarios with new heavy quarks and to extract 
the top-quark width within these scenarios. It can be applied to single top-quark measurements at the LHC. 

\section{Summary and Conclusions}

In this paper a case to measure the matrix element $|V_{ts}|$ from the top quark decays is demonstrated. 
In order to reduce the jet activity in top quark decays, it was suggested to tag the $W^\pm$ that decays 
leptonically, $W± \rightarrow \ell^\pm \nu_\ell$ $(\ell = e,\mu,\tau)$, and analyse the anticipated jet profiles 
in the signal process \tWs\ and the dominant background from the decay \tWb. A proposal to analyse 
the \V0 (\K0 and $\Lambda$) distributions in the s- and b-quark jets concentrating on the energy and 
transverse momentum distributions of these particles was presented. The \V0s emanating from the \tWb\ branch have 
displaced decay vertexes from the interaction point due to the weak decays $b \rightarrow c \rightarrow s$ 
and the b-quark jets are rich in charged leptons.
These distributions were used to train boosted decision trees, BDT(D). The BDT(D) 
response functions were obtained corresponding to the signal (\tWs) and background (\tWb). Detailed 
simulations undertaken with the Monte Carlo generator PYTHIA used to estimate the background 
rejection versus signal efficiency for three representative LHC energies $\sqrt{s} = $7, 10, and 14 \TeV .  A benchmark 
proved that with 10\% accuracy for the signal (\tWs) at a background (\tWb) rejection by a factor $10^3$ 
(required due to the anticipated value of the ratio $|V_{ts}|^2 / |V_{tb}|^2 \simeq 1.6 \times 10^{-3}$) can be achieved 
at the LHC $\sqrt{s}=14$ \TeV\ with an integrated luminosity of 10 \ifb\ .

In conclusion, a first study of its kind is presented here, showing that a direct measurement of $|V_{ts}|$ in 
top quark decays is feasible at the LHC. BDT results in typically 10\% efficiency for s-tagging with $10^3$ b-jet 
rejection. An oversimplified exercise for an integrated luminosity of 10 \ifb\ at $\sqrt{s} = 14$ \TeV\ 
taking $\sigma(t \bar{t}) \sim 1$ nb gave an estimated expected significance of $\sigma \sim 6$. 
A similar exercise for single top decays gave expected significance $\sigma \sim 3 $. Few alternative 
methods exist to measure directly $|V_{tq}|$ matrix elements and are presented briefly in this paper. 
The $|V_{ts}|$ method can complement these methods or directly be combined with them.

\end{document}